\documentclass[11pt]{article}
\usepackage{hyperref}
\usepackage{amssymb}
\usepackage{graphicx}
\usepackage{amsmath}
\usepackage{authblk}
\usepackage{cite}
\usepackage{physics}
\usepackage{revsymb}
\usepackage{float}
\usepackage{slashed}
\usepackage{tabularx}
\usepackage{breqn}
\usepackage{array}
\usepackage{verbatim}
\usepackage{dsfont}
\usepackage[section]{placeins}
\usepackage[a4paper, total={6.8in, 10in}]{geometry}
\numberwithin{equation}{section}
\newcommand{\be}{\begin{equation}}
\newcommand{\ee}{\end{equation}}
\newcommand{\bea}{\begin{eqnarray}}
\newcommand{\eea}{\end{eqnarray}}
\newcommand{\ba}{\begin{aligned}}
\newcommand{\ea}{\end{aligned}}

\begin{document}
\title{Maximal acceleration in Rainbow gravity}
\author{ {\bf {\normalsize Bhagya. R}\thanks{22phph03@uohyd.ac.in, r.bhagya1999@gmail.com},\,
{\bf {\normalsize Harsha Sreekumar}\thanks{22phph01@uohyd.ac.in, harshasreekumark@gmail.com}},\,
{\bf {\normalsize Suman Kumar Panja}\thanks{19phph17@uohyd.ac.in, sumanpanja19@gmail.com}}}\\
{\normalsize School of Physics, University of Hyderabad}\\{\normalsize Central University P.O, Hyderabad-500046, Telangana, India}\\ 
}
\date{}

\maketitle
\begin{abstract}
In this paper, we derive maximal acceleration of a massive particle in Rainbow gravity. Using eight-dimensional phase-space metric compatible with Rainbow gravity, we obtain the maximal acceleration, valid up to first order in the Rainbow gravity parameter $\eta$. Using the positivity condition on maximal acceleration, we find the upper bound on the Rainbow gravity parameter is of the order of $~10^{22}$ for positron and $10^{-44}$ for a black hole. After obtaining the expression for maximal acceleration for different choices of Rainbow functions, we derive corresponding modifications to Unruh temperature. Comparing with the observational value of the Unruh temperature, we find the upper bound on $\eta$ as $~10^{32}$ for positron radiation. 
We then derive geodesic equations for different choices of Rainbow functions and also obtain Newtonian limit of these geodesic equations. We find that the changes in the value of maximum acceleration, maximum temperature and Newtonian force equation are dependent on the choices of Rainbow functions.
\end{abstract}

\section{Introduction}

Quantum gravity seeks to unify two seemingly incompatible pillars of Physics–Einstein's general theory of relativity, which describes the force of gravity on cosmic scales, and quantum mechanics, which governs physics at the atomic scale and below. The common characteristic feature of all quantum gravity models such as loop quantum gravity, string theory, 
 non-commutative geometry 
  etc is the existence of an observer-independent fundamental length scale, which can be identified with the Planck length scale \cite{dsr}.
The existence of a fundamental length scale necessitates the modification of special theory of relativity. A modified relativity principle, admitting two fundamental constants, speed of light and Planck length had been constructed and is known as Doubly special relativity (DSR) \cite{dsr, kow}.
One well-studied model of gravity that was constructed on the framework of DSR is the Rainbow gravity \cite{lee, lee2, galan}. In Rainbow gravity (RG) paradigm the curvature of space-time near massive objects such as black holes causes different wavelengths of light to propagate at different speeds, leading to a ``Rainbow effect". Here the geometry of space-time gets affected by the energy of the test particle such that a single metric can no longer describe space-time at all energy scales but needs a family of metrics known as Rainbow of metrics. These metrics are parametrized by the ratio of the energy of the test particle($E$) and Planck energy($E_p$). 

The implications of RG on black holes physics 
\cite{ali, cycs, leiva, li, cunha, faizal, gim, panahiyan, silva, hamil} and their interactions with cosmic strings \cite{mota1, mota2} have been studied. In \cite{ali},  thermodynamical properties of Schwarzschild black hole in RG are studied. 
 A relation between black hole temperature and energy is also derived in the RG settings.
 In \cite{cunha}, a solution to the massive scalar field in RG modified Schwarzschild metric is obtained and using this the Hawking radiation is analyzed. In \cite{faizal}, charged dilatonic black hole thermodynamics is studied in RG. 
In \cite{mota2}, the behavior of scalar particles within the framework of RG, focusing on the space-time around a cosmic string under this modified gravity, is investigated. 
In \cite{mandal}, by incorporating the effects of the generalized uncertainty principle, the phase transition of a higher dimensional Schwarzschild black hole in the presence of RG is studied. In \cite{hendi}, the hydrostatic equilibrium equation of stars in (3+1) dimensional RG in the presence of cosmological constant is obtained by considering a spherically symmetric metric.

Various quantum gravity models propose the existence of a minimal length scale, which is linked to an upper limit on proper acceleration referred to as maximal acceleration \cite{Kothawala, nicolini, Torr1, Torr2}. In \cite{nicolini},  it is mentioned that maximal acceleration is related to the presence of a minimum time within a physical system, which can be intrinsic to the system or linked to a quantum gravity-induced ultraviolet cut-off. 
 In \cite{hlv}, within the framework of $\kappa$-deformed Minkowski space-time, the four-dimensional line element that accounts for causal connections is shown to impose a limit on proper acceleration. In \cite{vishnu}, maximal acceleration is studied in $\kappa$-space-time, and the non-commutative correction to maximal acceleration as well as the corresponding modified Unruh temperature is obtained. Similar studies have been done in DFR space-time \cite{suman}, demonstrating that non-commutativity decreases the value of maximal acceleration compared to the commutative case. Using Heisenberg's uncertainty principle, the upper bound on the acceleration of massive particles was derived in \cite{heis}. By analyzing the relationship between the temperature of vacuum radiation in a uniformly accelerated frame and the absolute maximum temperature of the radiation, maximal acceleration is obtained in \cite{brandt}. In \cite{papini}, the impact of maximal acceleration on compact stars is investigated and has revealed alterations in their stability condition. 
 Maximal acceleration has been utilized effectively in addressing UV divergences in local quantum field theory \cite{nfls}. The modified Unruh temperature was obtained from the Rindler metric embedded with maximal acceleration in \cite{BF}. Studies reported in \cite{kuwata, nuovo2} have explored upper limits on the mass of the Higgs boson by modifying Higgs-fermion interactions due to maximal acceleration. The existence of maximal acceleration has been found to modify the standard model of cosmology, leading to the avoidance of initial singularities and introducing an inflationary expansion \cite{gasp}. Thus, studying how RG will affect the maximal acceleration will have important implications. Here we obtain maximal acceleration in the presence of RG for different choices of Rainbow function. We construct the eight-dimensional phase-space metric using the modified dispersion relation and RG modified metric and derive modifications to maximal acceleration for three cases using this eight-dimensional RG phase space line element. We consider modification valid only up to the first order in the RG parameter. We show that in all three cases, the correction due to RG depends linearly on the mass of the test particle and differs only by numerical factors. Next, we derive the geodesic equations for RG. By taking the Newtonian limit of geodesic equations, we show that the RG-modified maximal acceleration leads to changes in the radial component of Newton's force equation.

This paper is organized in the following way. In sec. 2, we give a brief description of Rainbow gravity. In sec. 3, we derive the modified phase-space metric, valid up to the first order in $a$, where $a$ is $\frac{\eta}{E_p}$ and $\eta$ is the RG parameter. This is done by taking the direct sum of the energy-momentum relation and the space-time metric in the presence of RG. We derive an expression for the maximal acceleration, valid up to first-order in $a$, from the time-like line element defined in the eight-dimensional phase space. Using this we also derive the Unruh temperature. In sec. 4, we obtain the modified geodesic equation and find its Newtonian limit. In sec. 5, we summarise our results and give conclusions. 

\section{Rainbow Gravity}

In generalizing Galilean relativity to special theory of relativity one introduces a speed invariant scale. A similar approach is followed for constructing doubly special relativity, which contains an energy invariant scale, Planck scale \cite{dsr, kow} in addition to invariant speed. RG, constructed on the background of DSR theory has an energy-dependent metric \cite{BarceloVL, BarceloVL2005, Oriti2007, Gielen2013, Gielen2014, LafranceM1995, PengW2008, joao1}. 
RG refers to a unique aspect of this approach where space-time geometry is influenced by the energy of the test particle moving in it. This means that different observers using probes of varying energies will perceive different classical geometries. Consequently, if an inertial observer, freely falling through space-time utilizes a system of moving particles to investigate its geometry, the energy of these probe particles becomes a crucial factor. In essence, the metric describing space-time needs to incorporate the energy of the probe particles, resulting in a family of metrics known as ``Rainbow" metrics \cite{PengW2008}.
These metrics are parametrized by the ratio of the test particle's energy (E) to Planck energy ($E_p)$.
 
In RG, the effect of gravity on light depends on its wavelength. This would not be noticeable in low-gravity regions, but it becomes significant in regions of space-time where gravity is very strong such as near black holes. In RG, the modified dispersion relation of the test particle takes the form \cite{lee, lee2}
\be
E^2 f^2\Big(\frac{E}{E_p}\Big) - P^2 g^2\Big(\frac{E}{E_p}\Big) = m^2,  \label{Ma1}
\ee
where m is the mass and P is the 3-momentum of the test particle. Here $ f(\frac{E}{E_p})$ and $g(\frac{E}{E_p})$ are the Rainbow functions, where $E$ is the energy of the test particle, i.e., it is the energy scale at which background space-time is probed \cite{lee2} and $E_p$ is the Planck energy.
The Rainbow functions $f(\frac{E}{E_p})$ and $g(\frac{E}{E_p})$ satisfy the conditions,
\be
\lim_{{\frac{E}{E_p} \to 0}} f\Big(\frac{E}{E_p}\Big) = 1,
\lim_{{\frac{E}{E_p} \to 0}} g\Big(\frac{E}{E_p}\Big) = 1.
\ee
Transformation generated by non-linear representation of the Lorentz group leaves the above dispersion relation (\ref{Ma1}) invariant \cite{lee, lee2}. These transformations are generated by $K^{i}$ where
\be
K^i = U^{-1} L^i_0 U.
\ee
Here $L^i_0$ are the Lorentz generators and U is the map such that \cite{lee, lee2}
\be
U_a (E,p_i) = (U_0,U_i)= [E f(E),P_i g(E)]. \label{energy}
\ee
Also, generalization of this map acting on space-time coordinates gives \cite{joao1},
\be \label{T1}
U^a (x)= (U^0, U^i)=\Big(\frac{t}{f(E)}, \frac{x^i}{g(E)}\Big).
\ee
Using the above equation, one can write the metric as \cite{joao1},
\be \label{spacemetric}
d \hat{s}^2 = -\frac{dt^2}{f^2(E)} + \frac{dx_i^2}{g^2 (E)}.
\ee
Various choices of Rainbow functions motivated by physical considerations have been studied in \cite{lee, ellis, Jacob, Amel}.
In our study, we analyze the effect of three choices of Rainbow functions. The first choice is \cite{lee},
\be \label{rf1}
f(E) = g(E) = \frac{1}{1- \eta \Big(\frac{E}{E_p}\Big)} \approx 1 + a E,
\ee
where $\eta$ a dimensionless parameter of the model and $E_p$ is the Planck energy and $ a =\frac{\eta}{E_{p}}$. 
 The second choice is given by \cite{ellis},
\be \label{rf2}
f(E) = \frac{e^{\eta \frac{E}{E_{p}}}-1}{\eta \Big(\frac{E}{E_{p}}\Big)} \approx 1+\frac{a E}{2},~~~g(E) = 1.
\ee
The third choice for the RG function is given as \cite{Jacob, Amel}
\be \label{rf3}
f(E) = 1 ,~ g(E) =\sqrt{1 - \eta \Big(\frac{E}{E_p}\Big)^n}
 \approx 1 -\frac{a E}{2}.
 \ee
 Note that in all these cases we do the binomial expansion and consider only up to first order in $a$.
In our study for these three choices of Rainbow functions given in eq.(\ref{rf1}), eq.(\ref{rf2}) and eq.(\ref{rf3}), we derive the maximal acceleration. After calculating the maximal acceleration, we find the Unruh temperature associated with each of the above choices of Rainbow function. We also study the effects of maximal acceleration on the geodesic equation and its Newtonian limit.

\section{Rainbow gravity modified maximal acceleration}

In this section, we derive maximal acceleration of a particle moving under the influence of RG. In the Minkowski space time, maximal acceleration for a particle is derived from an 8-dimensional phase space line element which is constructed by taking the direct sum of the 4-dimensional flat space-metric and the energy-momentum relation. Then by applying the causality condition to the 8-dimensional line element, one obtains the maximal acceleration of the particle. Here we use the deformed energy-momentum relation given in RG to obtain the modified 8-dimensional phase space line element and derive the expression for maximal acceleration for a particle of mass m.

In our study, we consider three choices for Rainbow functions given in eq.(\ref{rf1}), eq.(\ref{rf2}) and eq.(\ref{rf3}) and find the maximal acceleration. Using the first choice of Rainbow function given in (\ref{rf1}) valid up to first order in $a$  we find,
\be \label{spacemetric1}
d \hat{s}^2 = -c^2\frac{dt^2}{f^2({E})} + \frac{dx_i^2}{g^2 (E)}
= -c^2 dt^2 (1- 2 a E)+ dx_i^2 (1- 2 a E).
\ee
From eq.(\ref{energy}), we find,
\be \label{E1}
d\hat{E}^2 = d{E}^2 + 4a E d E^2,~~d \hat{P}^2 = dP^2 + 2aE dP^2 +2a P_{i} dP^{i} dE,
\ee
where $i = 1, 2, 3$.
Using the direct sum of space-time metric given in eq.(\ref{spacemetric1}) and momentum-space metric obtained from eq.(\ref{E1}), we find the 8-dimensional phase-space line element as
\begin{multline} \label{PS1}
d \hat{S}^2 = -c^2 dt^2 \Big(1- 2a E\Big) + dx_i^2 \Big(1- 2a E\Big)+\frac{\hbar ^2}{\mu ^4 c^4}\bigg\{ -\frac{1}{c^2}\Big(d{E}^2 + 4aE d{E}^2\Big) +  d P^2\Big(1 + 2aE\Big)+ 2aP_{i} dP^{i} dE \bigg\}.
\end{multline}
Note that here $\frac{\hbar^2}{\mu^4 c^4}$ is introduced for dimensional reasons.
For time-like events, we have d$ \hat{S}^2  \leq 0$ and applying this causality condition in the above equation gives,
\be \label{timelike1}
-c^2 dt^2 \Big(1- 2aE\Big) + dx_i^2 \Big(1- 2aE\Big) +\frac{\hbar ^2}{\mu ^4 c^4}\bigg\{-\frac{1}{c^2}\Big(d{E}^2+ 4aE d{E}^2\Big) +  d P^2\Big(1 + 2aE\Big)+ 2aP_{i} dP^{i} dE \bigg\} \leq 0.
\ee
By dividing the above equation throughout by $ dt^2$, we find 
\begin{multline}
 -c^2 \Big(1- 2aE\Big) + \Big(\frac{dx_{i}}{dt}\Big)^2 \Big(1- 2aE\Big)+\frac{\hbar ^2}{\mu ^4 c^4}\bigg[-\frac{1}{c^2} \Big\{\Big(\frac{d{E}}{dt}\Big)^2 + 4aE \Big(\frac{d{E}}{dt}\Big)^2\Big\}\\ +  \Big(\frac{d P_i}{dt}\Big)^2\Big(1 + 2aE\Big)+ 2aP_{i} \frac{dP^{i}}{dt} \frac{dE}{dt}\bigg] \leq 0.
\end{multline}
Using eq.(\ref{rf1}) in the modified dispersion relation given in eq.(\ref{Ma1}) we find
\be \label{Dispersion1}
E^2 \Big(1 + aE\Big)^2 - P^2\Big(1 + aE \Big)^2= m^2,
\ee
valid up to first order in $a$. Differentiating the above equation with respect to time and keeping up to first order in $a$, we get
\be \label{derivative1}
\frac{d{E}}{dt} = \frac{P_{i}}{E} \frac{dP^{i}}{dt} \Big(1- a E +a \frac{P^2}{ E} \Big).
\ee
Using eq.($\ref{derivative1}$) in eq.($\ref{timelike1}$ ) and denoting $\frac{dx_i}{dt}$ as the velocity of the particle $v$, we obtain
\begin{multline}\label{final1}
  -(c^2-v^2)(1- 2a E) +\frac{\hbar ^2}{\mu ^4 c^4}  \Big(1+2aE +\frac{2 P^2 a}{E }\Big) \bigg\{ -\frac{1}{c^2} \frac{P^2}{E^2} \Big(\frac{dP_{i}}{dt}\Big)^2 + \Big(\frac{dP_{i}}{dt}\Big)^2\bigg\} \leq 0.
 \end{multline}
Here $P_i$ appearing in eq.(\ref{final1}) is the  relativistic 3-momenta in Minkowski space-time. 
Identifying the proper acceleration of the particle as $A$ we get, $\frac{dP_i}{dt}={mA}{(1-\frac{v^2}{c^2})^{-\frac{3}{2}}}$ and using this, we re-express the above equation as
\begin{multline}
  -(c^2-v^2)\Bigg\{(1- 2a E) -\frac{\hbar ^2 }{\mu ^4 c^4} \bigg(\frac{m^2A^2 c^6}{(c^2-v^2)^4}\bigg)\bigg(1+2aE +\frac{2 aP^2 }{E }\bigg) \bigg(1 -\frac{1}{c^2} \frac{P^2}{E^2}  \Bigg)\Bigg\} \leq 0.\label{finala}
 \end{multline}

As in the Minkowski space-time here also the maximum allowed speed is c and thus the acceleration becomes maximum when $v << c$.
For getting maximum acceleration, we consider an instantaneous rest frame of the particle where $v=0$, hence momentum $P$ is also zero. As the speed of the particle is bounded by c, when it is moving with speed c, it is expected to have zero acceleration. So for achieving maximum acceleration, the change in velocity should be maximum, which happens when velocity changes from a very low value to a very high value, say c.
Using these considerations in eq.($\ref{finala}$), we get
\be
-c^2 \Big(1- 2aE \Big) + \frac{\hbar ^2}{\mu ^4 c^4}\Big\{(m\hat{A}_{max})^2(1+ 2aE)\Big\}  \leq  0.\label{final1a}
\ee
After simplification of the above equation, we find the maximum acceleration satisfies
\be
\hat{A}_{max} \leq  \frac{mc^3}{\hbar}\bigg(1-2aE \bigg)
= \frac{mc^3}{\hbar}\bigg(1-2\frac{\eta mc^2}{E_p} \bigg).
\label{final1b}
\ee
Here we set the dimensionfull parameter $\mu$ as m which is the rest mass of the particle and $E=mc^2$.

One of the ingredients in the derivation of maximal acceleration in  \cite{cai1} is the postulate of metric measurability, which means the possibility to measure lengths with arbitrary precision. This assumption brings in with it, a length scale naturally associated with the massive particle whose acceleration one is studying. It is this length and mass scales through which $\hbar$ comes into the discussion of maximal acceleration (as the natural length scale associated with a massive particle is its Compton wavelength $\lambda= \frac{\hbar}{mc}$). The role of metric defined in the classical phase space is to enable the calculation of acceleration from kinematics alone. As pointed out in \cite{cai1}, in the limit $\hbar \rightarrow 0$, the configuration and momentum parts of the 8-dimensional phase space disconnect and we recover the classical result in this case also.

We next use this bound to obtain an upper cut-off on the temperature of thermal radiation. The temperature of thermal radiation, seen by an observer with uniform acceleration $A$, is given by the Unruh temperature, $T = \frac{\hbar A}{2 \pi k_B c}$ \cite{unruh}, where $k_B$ is the Bolzmann constant. So in this case, the maximum temperature associated with the thermal radiation is
\be \label{Tmax1}
T_{max} = 
\frac{mc^2}{2 \pi k_B } \bigg(1-2\frac{\eta E}{E_p}\bigg) = \frac{mc^2}{2 \pi k_B } \bigg(1-2\frac{\eta mc^2}{E_p}\bigg).
\ee
In obtaining the last equality, we have replaced the energy E of the probe particle by $mc^2$.

Now we consider the Rainbow functions given in eq.(\ref{rf2}). By taking the direct sum of space-time line element and energy-momentum relation, we obtain the expression for the 8-dimensional phase-space line element, valid up to first order in $a$ as
\be \label{8Dmetric2}
d\hat{S}^2= -c^2 dt^2 \Big( 1- a E\Big) + dx_i^2+ \frac{\hbar ^2}{\mu ^4 c^4} \bigg[-\frac{1}{c^2}\Big\{\Big(1+ 2a E\Big)dE^2 \Big\} + dP^2\bigg].
\ee
As in the previous case we impose the causality condition ($d\hat{S}^2 \leq 0$) and consider an instantaneous rest frame of the particle. Thus following the procedure discussed in the previous case, we get 

\be \label{amax2}
\hat{A}_{max}  
\leq \frac{mc^3}{\hbar} \bigg(1-\eta \frac{ E}{2 E_p}\bigg) = \frac{mc^3}{\hbar} \bigg(1-\eta \frac{ m c^2}{2 E_p}\bigg).
\ee 
Here again, the last equality is obtained by replacing E with $mc^2$.
Note that the correction terms in eq.(\ref{amax2}) and eq.(\ref{final1b}) differ by a multiplicative factor of 4.      By following the procedure discussed in the previous case we find the Unruh temperature corresponding to the value of maximal acceleration as
\be
T_{max} = \frac{mc^2}{2 \pi k_B } \bigg(1-\eta \frac{m c^2}{2 E_p}\bigg).\label{final2a}
\ee
The correction term differs by a factor of 4 from that in eq.(\ref{Tmax1}).
Corresponding to the third choice of Rainbow function given in eq.(\ref{rf3}) the 8-dimensional phase-space line element is obtained as,
\be
d\hat{S}^2 = -c^2 dt^2 +dx_i^2 (1+ aE)+\frac{\hbar ^2}{\mu ^4 c^4} \bigg\{-\frac{1}{c^2} dE^2 + dP^2 (1-a E)- 
a P_{i} dP^{i} dE \bigg\}.
\ee
Using this we find the maximal acceleration to satisfy
\be
A_{max} \leq \frac{mc^3}{\hbar}\Big( 1+ \eta \frac{ E}{2 E_p}\Big) = \frac{mc^3}{\hbar}\Big( 1+ \eta \frac{ mc^2}{2 E_p}\Big).\label{amax3}
\ee
Here we substituted $E = mc^2$ to obtain the last equality. As in the previous cases, here also using eq.(\ref{amax3}) we find the corresponding Unruh temperature as
\be
T_{max} = \frac{mc^2}{2 \pi k_B }\Big(1+\eta \frac{ mc^2}{2 E_p}\Big),
\ee
where we have taken $ E= mc^2$. 
We observe that here modifications due to RG increase the value of maximal Unruh temperature, unlike the first two choices of RG functions.

In all these three cases, we observe that the modification to the maximal acceleration depends, apart from the RG parameter $\eta$ on the mass of the test particle also. In the first two cases given in eq.(\ref{final1b}) and \ref{amax2}) we observe that the effect of RG is to reduce the maximal acceleration, while for the third choice, we find that $\hat{A}_{max}$ enhanced by RG. It was observed that for both  $\kappa$-deformed space-time \cite{ vishnu} and  DFR space-time \cite{suman}, the maximal acceleration gets modified in the presence of non-commutative space-time. In both these scenarios, the magnitude of maximal acceleration decreases due to non-commutativity of the space-time. In the limit  $\eta \rightarrow 0$, the above maximal acceleration expressions reduce to the result given in \cite{cai1, heis}. For a massless particle, $\hat{A}_{max}$ reduces to zero, as expected. Similar behavior is obtained in $\kappa$-space-time \cite{vishnu} and DFR space-time \cite{suman}. In the classical limit, i.e., $\hbar \rightarrow 0$, $\hat{A}_{max}$ diverges as it is observed in Minkowski space-time. By applying the condition that $\hat{A}_{max}$  should be positive in the first two choices of Rainbow functions (given in eq.(\ref{final1b}) and eq.(\ref{amax2})) and taking m as the positron mass, we find the upper bound on the RG parameter $\eta$ is of the order of $10^{22}$. By considering $m$ as mass of a black-hole (m= $10 ^{6}M_{\odot}$) for the first two choices of Rainbow functions, the upper bound on the value of $\eta$ is obtained as of the order of $10^{-44}$. We observe that in eq.(\ref{amax3}), the correction term to the maximal acceleration is positive as long as $\eta > 0$.

From the experimental results on the study of Unruh temperature associated with positron radiation \cite{cohen}, the upper bound on the value of $\eta$ is obtained to be of the order of $10^{32}$. And from the analysis done on the Unruh temperature associated with black-hole radiation, it is found that $\eta$ is of the order of $10^{-100}$.

\section{Rainbow gravity modified geodesic equation and its Newtonian limit}

In this section, we derive the geodesic equation in the presence of RG and obtain the corresponding Newtonian limit. We do this for all three choices of Rainbow function. The general form of the line element in the RG is \cite{joao1}
\be \label{gmetric}
d\hat{s}^2 = \hat{g}_{0 0} d\hat{x}^{0}d\hat{x}^{0} + \hat{g}_{ij} d\hat{x}^{i}d\hat{x}^{j},
\ee
where $\hat{g}_{00}=\frac{g_{00}}{f^{2}(E)}$ and $\hat{g}_{ii}=\frac{g_{ii}}{g^{2}(E)}$. Using eq.(\ref{rf1}) and eq.(\ref{final1b}), by taking $\hbar = c=1$, we obtain an expression for RG parameter as,
\be \label{lambda1}
a \leq \frac{1}{2m} \Big(1- \frac{\hat{A}_{max}}{m}\Big).
\ee
Using eq.(\ref{lambda1}) and eq.(\ref{spacemetric}) we find the line element in eq.(\ref{gmetric}) as
\be
d\hat{s}^2 = g_{00} \bigg\{1-\frac{E}{m}(1- \frac{A_{max}}{m})\bigg\}dx^0 dx^0 + g_{ij}\bigg\{(1-\frac{E}{m}(1- \frac{A_{max}}{m})\bigg\}dx^i dx^j.
\ee
Using this metric, we construct the modified geodesic equation in the presence of RG. For this, first we evaluate the deformed Christoffel symbols  $\hat{\Gamma}^{a}_{\mu \nu} = \frac{1}{2} \hat{g}^{a \alpha}[\partial_\mu \hat{g}^{\alpha \nu} + \partial_\nu \hat{g}^{\alpha \mu} - \partial_\alpha \hat{g}^{\mu \nu} ]$. 
We find the non-vanishing Christoffel symbols as
\bea
&\hat{\Gamma}^{0}_{00} = \frac{1}{2} \hat{g}^{00} \partial_{0} \hat{g}_{00};~~
\hat{\Gamma}^{0}_{i0} =\hat{\Gamma}^{0}_{0i}= \frac{1}{2} \hat{g}^{00} \partial_{i} \hat{g}_{00};~~
\hat{\Gamma}^{0}_{ij}=\frac{1}{2} \frac{f^2(E)}{g^2(E)} \hat{g}^{00} \partial_{0} \hat{g}_{ij};~~ 
\hat{\Gamma}^{i}_{00}=\frac{1}{2} \frac{g^2(E)}{f^2(E)} \hat{g}^{ij} \partial_{j} \hat{g}_{00};& \nonumber\\
&\hat{\Gamma}^{i}_{0j} =\hat{\Gamma}^{i}_{j0}=\frac{1}{2}  \hat{g}^{ik} \partial_{0} \hat{g}_{kj};~~
\hat{\Gamma}^{i}_{jk}=\frac{1}{2} \hat{g}^{im}[ \partial_{j} \hat{g}_{mk} +\partial_{k} \hat{g}_{mj} - \partial_{m} \hat{g}_{jk}].& \label{cris}
\eea
The modified geodesic equation in the RG is given as $\frac{d^2\hat{x}^{\mu}}{d\tau^2} + \hat{\Gamma}^{\mu}_{\nu a} \frac{d\hat{x^{\nu}}}{d \tau} \frac{d\hat{x^{a}}}{d \tau} =0
$.
Using eq.(\ref{T1}) and  eq.(\ref{cris}), we find the geodesic equation for $\mu = 0$ to be,
 \be \label{01}
  (1-a E)\frac{d^2{x}^{0}}{d\tau^2}+ (1-2 a E )\frac{1}{2} g^{00} \partial_{0}g_{00} \frac{d{x^{0}}}{d \tau} \frac{d{x^{0}}}{d \tau} +  (1-2a E ) g^{00}\partial_{i}g_{00}\frac{d{x^{i}}}{d \tau} \frac{d{x^{0}}}{d \tau}+ \frac{1}{2} (1-2a E)g^{00} \partial_{0}g_{ij}\frac{d{x^{i}}}{d \tau} \frac{d{x^{j}}}{d \tau}=0.
 \ee
Similarly we find geodesic equation for $\mu = i$ to be
\begin{multline} \label{i1}
(1-a E)\frac{d^2{x}^{i}}{d\tau^2}+ \frac{1}{2}(1-2 a E ) g^{ij} \partial_{j}g_{00} \frac{d{x^{0}}}{d \tau} \frac{d{x^{0}}}{d \tau} + (1-2a E ) g^{ik}\partial_{0}g_{kj}\frac{d{x^{0}}}{d \tau} \frac{d{x^{j}}}{d \tau}\\+ \frac{1}{2} (1-2a E)g^{im}( \partial_{j}g_{mk}+\partial_{k}g_{mj}-\partial_{m}g_{jk})\frac{d{x^{j}}}{d \tau} \frac{d{x^{k}}}{d \tau}=0.
\end{multline}
The Newtonian limit of the above equation is derived by imposing conditions, (i) particle is moving slowly, (ii) gravitational field is static and (iii) as well as it is weak. These conditions translate into
\be
 \frac{d{x}^i}{d\tau}<<\frac{d{x}^{0}}{d\tau},\label{wfa1}
\ee
\be
 \frac{\partial {g}_{\mu\nu}}{\partial t}=0,\label{wfa2}
\ee
and
\be\label{wfa}
 {g}_{\mu\nu}={\eta}_{\mu\nu}+{h}_{\mu\nu},~~~|{h}_{\mu\nu}|<<1,
\ee
respectively. Using equation given in  eq.(\ref{lambda1}) in eq.(\ref{01}) and using the above conditions we find,
\be
\frac{d^2{x}^{0}}{d\tau^2} \bigg\{(1-\frac{E}{2m}(1-\frac{\hat{A}_{max}}{m})\bigg\} =0,
\ee 
which implies, 
\be
\frac{d^2{x}^{0}}{d\tau^2} =0~~\text{and}~~\frac{d x^{0}}{d \tau}=constant.\label{con}
\ee
Next using the conditions given in eq.(\ref{wfa1}) to eq.(\ref{wfa}) with eq.(\ref{lambda1}) in eq.(\ref{i1}) we  find
\be
\frac{d^2{x}^{i}}{d\tau^2} \bigg\{1-\frac{E}{2m}\Big(1-\frac{\hat{A}_{max}}{m}\Big)\bigg\} + \frac{1}{2} \bigg\{1-\frac{E}{m}\Big(1-\frac{\hat{A}_{max}}{m}\Big)\bigg\}\frac{dx^0}{d\tau }\frac{dx^0}{d\tau}\bigtriangledown_i h_{00} =0. \label{wfb}
\ee
Using the $\frac{d^2 x_i}{d \tau^2} = \frac{d^2 x_{i}}{dt^2} \Big(\frac{dx_0}{d\tau}\Big)^2$ and eq.(\ref{con}) we rewrite the above equation as
\be
\frac{d^2x^i}{dt^2} \bigg\{1-\frac{E}{2m}\Big(1-\frac{\hat{A}_{max}}{m}\Big)\bigg\} + \frac{1}{2} \bigg\{1-\frac{E}{m}\Big(1-\frac{\hat{A}_{max}}{m}\Big)\bigg\}\bigtriangledown_i h_{00} =0. \label{wfb1}
\ee
Next we multiply the above equation by $[1+\frac{E}{2m}(1-\frac{\hat{A}_{max}}{m})]$ and keeping terms up to first order in $a$, obtain
\be \label{h1}
\frac{d^2x^i}{dt^2}  + \frac{1}{2} \bigg\{1-\frac{E}{2m}\Big(1-\frac{\hat{A}_{max}}{m}\Big)\bigg\} \bigtriangledown_i h_{00} =0.
\ee
From the above expression, we find that $\hat{h}_{00}=\bigg\{1-\frac{E}{2m}\Big(1-\frac{\hat{A}_{max}}{m}\Big)\bigg\} h_{00} $.
where $h_{00} = -\frac{2M}{r}$. Using this in eq.(\ref{h1}) and comparing this with the deformed Newton's equation, $F^i=m \frac{d^2 \hat{x}^{i}}{dt^2}$, we find
\be \label{F1}
\hat{F}_i=F_i\bigg\{1-\frac{E}{2m}\big(1-\frac{\hat{A}_{max}}{m}\big)\bigg\}.
\ee
Here, $F_i = -\frac{mM}{r^2}$, is the Newtonian force in flat space-time. It is interesting to note that in the above equation, the contribution from RG only affects the radial component as in \cite{hjm}. This observation aligns with findings in $\kappa$-defomed space-time \cite{vishnu} and in DFR space-time \cite{suman}. We observe that our result indicates that the modification of the Newtonian force depends on $a$, as well as on the energy of the particle $E$.

For the choice of Rainbow function given in eq.(\ref{rf1}), we have obtained an expression for $\hat{A}_{max}$ in eq.(\ref{amax2}). Putting $\hbar = c=1$ in eq.(\ref{amax2}) we obtain an expression for RG parameter as,
\be \label{lambda2}
a \leq \frac{2}{m} \Big(1- \frac{\hat{A}_{max}}{m}\Big).
\ee
Note that eq.(\ref{lambda1}) and eq.(\ref{lambda2}) differ by a multiplicative factor 4. By following the same procedure as in the previous case, we find the following geodesic equations
 \begin{multline} \label{02}
 (1-\frac{a E}{2})\frac{d^2{x}^{0}}{d\tau^2} + \frac{1}{2} (1-a E ) g^{00} \partial_{0}g_{00} \frac{d{x^{0}}}{d \tau} \frac{d{x^{0}}}{d \tau} +  (1-\frac{a E}{2} ) g^{00}\partial_{i}g_{00}\frac{d{x^{i}}}{d \tau} \frac{d{x^{0}}}{d \tau}+ \frac{1}{2} (1+a E)g^{00} \partial_{0}g_{ij}\frac{d{x^{i}}}{d \tau} \frac{d{x^{j}}}{d \tau}=0,
 \end{multline}
and
\begin{multline} \label{i2}
\frac{d^2{x}^{i}}{d\tau^2} + \frac{1}{2}(1-2 a E ) g^{ij} \partial_{j}g_{00} \frac{d{x^{0}}}{d \tau} \frac{d{x^{0}}}{d \tau} +  (1-\frac{a E}{2} ) g^{ik}\partial_{0}g_{kj}\frac{d{x^{0}}}{d \tau} \frac{d{x^{j}}}{d \tau}\\+ \frac{1}{2} g^{im}[ \partial_{j}g_{mk}+\partial_{k}g_{mj}-\partial_{m}g_{jk}]\frac{d{x^{j}}}{d \tau} \frac{d{x^{k}}}{d \tau}=0.
\end{multline}
The Newtonian limit of the eq.(\ref{02}) is
\be
\frac{d^2{x}^{0}}{d\tau^2} \bigg\{1- \frac{E}{m}\Big(1-\frac{\hat{A}_{max}}{m}\Big)\bigg\} =0,\label{ca2}
\ee
where we have used eq.(\ref{lambda2}) for $a$. Here also this implies, 
$\frac{d^2{x}^{0}}{d\tau^2} =0,~\frac{d x^{0}}{d \tau}=constant.\label{ca2a}
$
And from eq.(\ref{i2}) we obtain
\be
\frac{d^2{x}^{i}}{d\tau^2} + \frac{1}{2} \bigg\{1-\frac{4E}{m}\Big(1-\frac{\hat{A}_{max}}{m}\Big)\bigg\}\frac{dx^0}{d\tau }\frac{dx^0}{d\tau}\nabla_i h_{00} =0. \label{ca2b}
\ee
Using $\frac{dx^0}{d\tau}$ is a constant in the above eq. (\ref{ca2b}) we find
\be \label{h2}
\frac{d^2x^i}{dt^2} + \frac{1}{2} \bigg\{1-\frac{4E}{m}\Big(1-\frac{\hat{A}_{max}}{m}\Big)\bigg\}\nabla^i h_{00} =0.
\ee
From the above expression, we define $\hat{h}_{00}=[1-\frac{4E}{m}(1-\frac{\hat{A}_{max}}{m})] h_{00} $,
where $h_{00} = -\frac{2M}{r}$. Using this in eq.(\ref{h2}) and comparing this with the modified Newton's equation, $F^i=m \frac{d^2 \hat{x}^{i}}{dt^2}$, we find
\be \label{F2}
\hat{F}_i=F_i\bigg\{1-\frac{4E}{m}\Big(1-\frac{\hat{A}_{max}}{m}\Big)\bigg\},
\ee
$F_i = -\frac{mM}{r^2}$ represents Newton's force equation in the Minkowski space-time. It is evident that the influence of RG on the force is solely in the radial component. As we have observed in eq.(\ref{F1}), here also the RG effect reduces the absolute value of the Newtonian force. Similar features were reported in \cite{suman, vishnu}. It is important to note that the modification in the Newtonian force is dependent upon both the RG parameter $\eta$ and E, energy of the particle probing the background space-time.

Finally for the choice of Rainbow function given in eq.(\ref{rf3}), from eq. (\ref{amax3}) we find the parameter $a$ as (when we set $\hbar$ = 1= c) satisfying
\be \label{lambda3}
a \geq \frac{2}{m} \Big(\frac{\hat{A}_{max}}{m} -1\Big).
\ee
As in the earlier cases, here also we find the geodesic equations as
 \be \label{03}
 \frac{d^2{x}^{0}}{d\tau^2} + \frac{1}{2} g^{00} \partial_{0}g_{00} \frac{d{x^{0}}}{d \tau} \frac{d{x^{0}}}{d \tau} +  (1+\frac{a E}{2} ) g^{00}\partial_{i}g_{00}\frac{d{x^{i}}}{d \tau} \frac{d{x^{0}}}{d \tau}+ \frac{1}{2} (1+\frac{3a E}{2})g^{00} \partial_{0}g_{ij}\frac{d{x^{i}}}{d \tau} \frac{d{x^{j}}}{d \tau}=0.
 \ee
 and
\begin{multline} \label{i3}
(1+\frac{a E}{2})\frac{d^2{x}^{i}}{d\tau^2} + \frac{1}{2}(1- a E ) g^{ij} \partial_{j}g_{00} \frac{d{x^{0}}}{d \tau} \frac{d{x^{0}}}{d \tau} +  (1+\frac{a E}{2} ) g^{ik}\partial_{0}g_{kj}\frac{d{x^{0}}}{d \tau} \frac{d{x^{j}}}{d \tau}+ \\(1+a E)\frac{1}{2} g^{im}[ \partial_{j}g_{mk}+\partial_{k}g_{mj}-\partial_{m}g_{jk}]\frac{d{x^{j}}}{d \tau} \frac{d{x^{k}}}{d \tau}=0.
\end{multline}
For deriving the Newtonian limit, we apply the three conditions (see eq.(\ref{wfa1}) to eq.(\ref{wfa})) on eq.(\ref{03}) and find 
$\frac{d x^{0}}{d \tau}=\text{constant}$.
Thus, the Newtonian limit to eq.(\ref{i3}) is
\be
\Big(1+\frac{a E}{2}\Big)\frac{d^2{x}^{i}}{d\tau^2} + \frac{1}{2}\Big(1- a E \Big) \frac{dx^0}{d\tau }\frac{dx^0}{d\tau}\nabla_i h_{00} =0. \label{ca3a}
\ee
We rewrite the above equation as
\be \label{h3}
\Big(1+\frac{a E}{2}\Big)\frac{d^2x^i}{dt^2} + \frac{1}{2} \bigg(1-a E\bigg)\nabla^i h_{00} =0.
\ee
Now multiply the above equation throughout by $(1-\frac{a E}{2})$ and using eq.(\ref{lambda3}) gives,
\be
\frac{d^2x^i}{dt^2} + \frac{1}{2} \bigg\{1-3\frac{E}{m}\Big(\frac{\hat{A}_{max}}{m}-1\Big)\bigg\}\nabla^i h_{00} =0.
\ee
From the above expression, we find that $\hat{h}_{00}=\Big(1+\frac{3E}{m}(1-\frac{\hat{A}_{max}}{m})\Big)h_{00} $.
where $h_{00} = -\frac{2M}{r}$. Thus by comparing this with the deformed Newton's equation, $F^i=m \frac{d^2 \hat{x}^{i}}{dt^2}$, we find
\be \label{F2}
\hat{F}_i=F_i\bigg\{1+\frac{3E}{m}\Big(1-\frac{\hat{A}_{max}}{m}\Big)\bigg\}.
\ee
Here $F_i=-\frac{mM}{r^2}$ is Newton's force equation. 
We observe in this case too contribution from RG to the Newtonian force has only radial component. We observe that here correction term due to RG enhances the force, unlike in the previous two cases.
In all three cases, we have taken modifications to the force expression up to first order in $a$. In the limit, $a \rightarrow 0$ all the modified force expression obtained in three cases reduce to the usual expression of Newton's force.

\section{Conclusion}

In our study, we have derived the maximum possible acceleration of a massive particle in the RG. For this, we have constructed the modified 8-dimensional  phase space line element, by taking the direct sum of RG modified flat space metric and dispersion relation in RG, valid up to first-order in the parameter $a$. Ensuring causal connection for the 8-dimensional phase-space line element, we have derived the RG modified maximal acceleration, for three different choices of Rainbow function. This correction relies on the mass of the test particle and vanishes for massless particles. For all three cases, in the limit, $a \rightarrow 0$, RG modified maximal acceleration reduce to the usual result \cite{cai1, heis}. We then use this upper bound on proper acceleration in Unruh temperature formula and discuss the bound on Unruh temperature. We observe that in first two cases, RG modifications reduce the maximal acceleration and maximal temperature. But in the third case, we note that modification enhances the maximal acceleration and maximal temperature. In the classical limit (as $\hbar \rightarrow 0 $) maximal acceleration becomes infinity.

By imposing the requirement that $\hat{A}_{max}$ must be positive and assuming the mass $m$ corresponds to that of a positron, we determine that the upper limit on the RG $\eta$ is on the order of $10^{22}$ for the first two choices of RG function. However, if we take $m$ to be the mass of a black hole (m= $10 ^{6}M{\odot}$), the maximum allowed value for $\eta$ is of the order of the  $10^{-44}$. Then using the modified Unruh temperature for each choice of Rainbow function with the observational result \cite{cohen}, we found the upper bound on the $\eta$ as $~10^{32}$ for positron radiation and of the order of $10^{-100}$ for black hole radiation.

Utilizing the RG-modified metric, we have calculated the deformed Christoffel symbol up to first-order in $a$ and derived the RG-modified geodesic equations. We express the modifications in the geodesic equations in terms of maximal acceleration. From these modified geodesic equations we have obtained the corresponding Newtonian limit in RG. Results show that the correction term in the modified  Newton’s force equation is solely radial, similar to the results obtained in \cite{vishnu, suman}. We observe that modifications to Newton's force in first two cases decrease its magnitude but in third case correction term increases the magnitude of Newton's force.  In all cases, we have taken modification valid up to first order in  $a$. All results do match with the well-known result in flat space-time, that is, in the limit $a \rightarrow 0$.

\section{Acknowledgement}

We thank E. Harikumar for useful discussions and comments. BR thanks DST-INSPIRE for support through the INSPIRE fellowship (IF220179). HS thanks Prime Minister Research Fellowship (PMRF id:3703690) for the financial support. S.K.P thanks UGC, India, for the support through the JRF scheme (id.191620059604).

\end{document}